# Evolutionary Dynamics of Complex Networks: Theory, Methods and Applications

**Alireza Abbasi**
School of Engineering & IT,
University of New South Wales Canberra
ACT 2600, Australia

**Liaquat Hossain (Corresponding Author)**
Liaquat Hossain
Professor, Information Management
Division of Information and Technology Studies
The University of Hong Kong
lhossain@hku.hk

Honorary Professor, Complex Systems
School of Civil Engineering
Faculty of Engineering and IT
The University of Sydney, Australia
Liaquat.hossain@sydney.edu.au

**Rolf T Wigand**
Maulden-Entergy Chair & Distinguished Professor
Departments of Information Science & Management
UALR, 548 EIT Building
2801 South University Avenue
Little Rock, AR 72204-1099, USA

# Evolutionary Dynamics of Complex Networks: Theory, Methods and Applications

**Abstract**

We propose a new direction to understanding evolutionary dynamics of complex networks. We focus on two different types of collaboration networks—(i) academic collaboration networks (co-authorship); and, (ii) disaster collaboration networks. Some studies used collaboration networks to study network dynamics (Barabási & Albert, 1999; Barabási et al., 2002; Newman, 2001) to reveal the existence of specific network topologies (structure) and preferential attachment as a structuring factor (Milojevi 2010). The academic co-authorship network represents a prototype of complex evolving networks (Barabási et al., 2002). Moreover, a disaster collaboration network presents a highly dynamic complex network, which evolves over time during the response and recovery phases of a disaster. We have chosen these two networks even though they are different in terms of complexity and domain but have common properties: open-based communication platform (self-organized system); geographically disperse actors (locational and cultural diversity of actors); autonomous behavior systems (non- hierarchy structure). The results show that the academic collaboration network has all the properties of small-world networks and can be considered a real small-world network with the result of its structural analysis maybe being extendable for other real-networks who share the common grounds of small-world properties. Our results further illustrate that the networks' clustering coefficients remain almost static over the evolution of both networks while the network is expanding over time through adding new actors and links to the networks.

*Keywords: Evolutionary Dynamics; Complex Networks; self-organized systems; autonomous behavior; network theory; social networks analysis; collaborative systems.*

# Evolutionary Dynamics of Complex Networks: Theory, Methods and Applications

**Introduction**

Collaboration is one of the defining features of modern science (Milojevi 2010). It is defined as "working together for a common goal and sharing of knowledge" (Hara, Solomon, Kim, & Sonnenwald, 2003) (p. 853). Therefore, collaboration can be regarded as a social process which enables the collective shared intelligence of a group (Bordons & Gómez, 2000; Milojevi 2010; Shrum, Genuth, & Chompalov, 2007). As Milojevic (2010, p. 1410) formulated, "The most commonly used methods for studying collaboration networks have been: bibliometrics (Bordons & Gómez, 2000); social network analysis/network science (Barabási et al., 2002; Kretschmer, 1997; Newman, 2001c, 2004a; Wagner, 2008; Wagner & Leydesdorff, 2005); qualitative methods of observation and interviews (Hara et al., 2003; Shrum et al., 2007); and surveys.

Currently, however, it is not clear which collaboration data is useful for evaluating the academic community. Although there is a large set of potential collaboration data (e.g., joined conference organization, joined research proposal submissions, joined publications, joined conference attendance, and teacher-student relationships) (Abbasi, Altmann, & Hossain, 2011), which qualifies for being analyzed through appropriate network measures but co-authorship is the most visible indicator of scientific collaboration and has thus been frequently used to measure collaborative activity, especially in bibliometric and network analysis studies. Bibliometric studies of co-authorship have emphasized the effects of collaboration on scientific activity as well as on organizational and institutional aspects of collaboration applying different units of analysis (i.e., authors, institutions, and countries) (Milojevi 2010).

Since previous reports showed, worldwide, losses from disasters related to natural hazards had risen dramatically (Munich Re Group, 2005; Swiss Reinsurance Company, 2006) and that trend was growing due to human-induced climate change (Crompton & McAneney, 2008); so it is expected to grow exponentially now and in the future. A disaster is defined as a destructive event creating negative social and economic conditions that interrupt the day-to-day activities of a society (Abbasi, Hossain, Hamra, & Owen, 2010; Hossain & Kuti, 2010; Kapucu, 2005; Waugh, 2003). In such special events, as the situations become more severe and complex, a coordination network consisting of different organizations and agencies is needed to manage and respond to the events optimally. Therefore, it is important for organizations to form cooperative information exchanges in order to respond to disasters effectively and efficiently to achieve a common goal (Abbasi & Kapucu, 2012). Poor coordination is a fundamental problem for an emergency response, and is primarily due to lack of good communication between and among organizations; a lack of up-to-date and relevant information circulating through the emergency response network; and, insufficient access to data and action plans (Hossain & Kuti, 2010; Van Borkulo, Scholten, Zllatanova, & van den Brink, 2005).

Coordinating activities or tasks in a complex system during a disaster for effective response is one of the most important issues to protect human, natural lives as well as protecting these from infrastructure damage. In any system and especially disasters, if the situation becomes more complex, by increasing interactions and interdependency among actors (e.g., individuals and organisations), the need for coordination increases in order to achieve a common goal. It has been documented in previous studies that the efficiency of disaster response is influenced by the severity of disaster, type and amount of resources available, number of jurisdictions involved, and complexity of the response strategies (Comfort, Ko, & Zagorecki, 2004). On the other hand, flexibility gives the involved participants and organizations ability to respond to unanticipated events. During an emergency situation, flexibility will support organizations to be properly prepared and improvise to fit the requirements of the current situation (Mendonça, Beroggi, & Wallace,

2001). Thus, flexibility and adaptation of the coordinating networks in uncertain and dynamic conditions (e.g., disasters) needs to be seen pivotal to effective response and recovery efforts (Dwyer & Owen, 2009).

Complex adaptive systems exhibit coherence and persistence in the face of change and dependence on extensive interaction (Holland, 1993). Crisis environments are complex, dynamic and place lives, property, and continuity of operations in our communities at risk. Concepts from the emerging science of complex adaptive systems (CAS) help us to understand these environments better and offer fresh insight into the challenging problems of inter-organizational coordination (Comfort, 1999). Modern organizational environments are becoming more complex at an increasing rate, largely through networks (Kauffman, 1993; Scott, 1998; Waldrop, 1992; Weick, 2001). "The interactions of organizations in a large system can generate greater complexity than the organizations themselves. Today's organizations have become too interconnected and too complex to be managed by traditional top-down and hierarchical means" (Kapucu, 2004). The science of complex adaptive systems offers a new mind-set and a scientific basis for managing complex partnerships using the principles of self-adaptive systems (Clippinger, 1999).

**Common properties of the two networks**

*Open-based communication platform (self-organized system)*

Disaster collaboration networks resemble open-based communication platforms. Although usually there is a systematic hierarchy for response to disasters, they follow a bottom-up perspective. A disaster collaboration network is a self-organized system meaning that individual agents (e.g., personnel, agencies and organizations) organize their communication with other involved actors as an incident (disaster) evolves. Similarly, an academic collaboration (co-authorship) network does not have a hierarchical structure and individual agents (i.e. authors) organize whole network topology and state through their collaboration with others. In disaster collaboration networks, bottom up information could be used at the top down command and control in ensuring adequacy in situation awareness of the affected communities. In these networks, in order to transmit the local knowledge of the affected areas/communities to central command and control so that a realistic disaster response and recovery effort should become deliverable.

### *Geographically dispersed groups (locational and cultural diversity of actors)*

During response to a complex disaster, several agents from different organizational cultures and locations cooperate with each other for a common goal. For instance, in order to respond to a large bushfire several agencies from different organizational cultures (such as fire services, police, ambulance) and usually from different jurisdictions come together. Similarly, scholars from different locations around the world and diverse disciplines collaborate with each other to have a joint publication.

### *Autonomous behaviour systems (non-hierarchical structure)*

There is no hierarchical structure (command and control) among authors in an academic co-authorship collaboration network meaning that often the authors are not forced to collaboration with another author. It is usually the same in disaster collaboration networks too. Although most emergency response organizations plan their general tasks and probable contacts, but as the disaster circumstances are very dynamic their links shape autonomously. In addition, academic collaboration networks can be referred to as stable networks while disaster collaboration networks are more complex and dependent on the environment. Analysis of these two types of networks would provide us with new insights about the effects of complexity on the behavior of network dynamics and evolution.

## Towards a Complex Adaptive Systems Cycle Model

In our proposed model presented in Figure 1, the co-evolution of local dynamics versus global dynamics of networks over time has been explored. To study dynamics *on* networks, the nodes' local structure correlation to networks topology in each time slot (static analysis) has been examined. On the other hand, as networks evolve over time represents that new nodes and links add to the network and both nodes and the whole network (locally and globally) structure changes due to different link attachment rules over the evolution of the network. To study dynamics of network over time, the dynamics of both node's local structure and network's topology compare to its immediate previous time slot has been further explored.

Using different dimensions of our model framework discussed above, we propose the following research questions in order to investigate collaboration networks' co-evolutionary adaptive behaviour formulated by (Gross & Blasius, 2008): *dynamics of networks* (the topology of the network changes in time according to specific, often local, rules); and *dynamics on networks* (the topology of the network remains static while the states of the nodes change dynamically).

- Do positions (roles) of existing actors in their collaboration network associate with their future attachment behavior?
- If so, what is the most favorable position to attract more partners?
- Does this phenomenon differ between a stable collaboration network and a dynamic complex collaboration network?

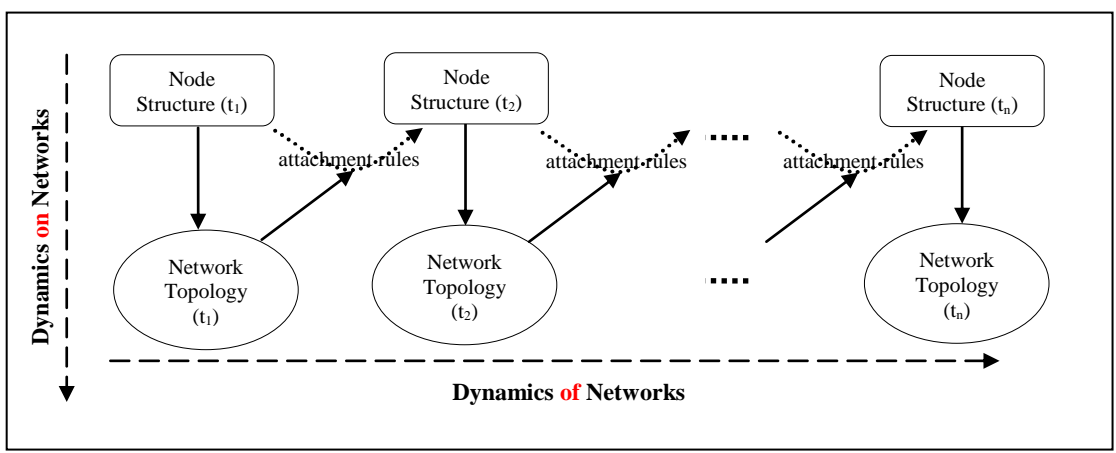

**Figure 1: A co-evolutionary dynamic model**

Providing answers for these questions helps us to understand the selection mechanisms of network evolution during the evolution of networks. Such understanding can help researchers, decision makers and practitioners to reach their goals. In particular the findings could help to identify the behavioral attachment patterns particularly identifying which characteristic(s) of existing actors more attract new actors or cause new actors attach to them (to evaluate the selection mechanism). Thus the following questions arise:

- What are the logics of actors for their attachment behavior during the evolution of a network? (How do actors choose links during their collaboration network evolution?)
- Do the complexity and dynamicity of the collaboration networks affect the actors' logics of attachment?
- Which attachment logic is the best driver of network evolution in each case of (stable or dynamic) the collaboration network?

Answering these questions helps us to check if different attachment logics (e.g., homophily, preferential attachment, embedding, and multi-connectivity) change network topologies. And if they do, what is the effect of each one on the network structural change. The findings help to determine how the topology of the network changes in time according to actors' attachment logics, which is useful in investigating the retention mechanism by analyzing dynamics of networks. It further assists in finding attributes of the network topology which remains static while the states of the actors change over time representative by evaluating the variation mechanism by investigating dynamics on networks.

**Social Networks Research Design, Data Collection and Analysis**

In order to answer these research questions, we extract two types of collaboration networks (academic and disaster) from the datasets, which have been normalized or our analysis. Then, we apply social network analysis metrics to explore each and all network properties. Social network metrics also help us to identify the parameters of our model. Subsequently, we use statistical analysis to examine our hypotheses.

Network data is most commonly collected on relationships linking a set of social units on a single type of relationship and is most often represented in a matrix X={$x_{i,j}$} where $x_{i,j}$ is the strength from unit i to unit j (Papachristos, 2006; Wigand, 1988). Data on social networks can be collected for all links (whole network data) or for the sets of links surrounding a specified social entity (ego-network data) (Marsden, 1990). The socio-centric network approach identifies structural patterns of interaction in a network but the ego-centric network approach helps to understand the social structure of the ego (Abbasi, Chung, & Hossain, 2012; Chung, 2009). Network data can come from "questionnaires, archives, observations, diaries, electronic traces and experiments" (Marsden, 1990). Survey and questionnaires are the most widely used method followed by archival resources (Marsden, 1990). In surveys (questionnaires) respondents are usually asked to identify their contacts for a specific type of relationship. Thus respondents self-report their network. Archive resources could from social network systems (e.g., Facebook, Linkedin, google+), online games, forums in which the people's interaction log shows who is connecting to whom. Also, text reports of interactions especially during an emergency response event such as newspaper reports on the responding organizations to disasters (e.g., Hurricanes in USA, Fire in Australia) are other types of archival resources for which some text mining techniques are needed to extract network data.

*Academic Collaboration Network Data*

In an academic collaboration network, nodes are referred as authors and ties (links) are co-authorship relations among them. A tie exists between each two authors (scholars) if they have at least one co-authored publication. In general, scientific collaboration (co-authorship) networks can be represented as a graph. The nodes (actors, vertices) of the graph represent authors and the links (ties, edges) between each two nodes indicate a co-authorship relationship between them. The weights of links denote the number of publications that two authors (co-authors) have jointly published.

Academic collaboration networks are complex types of social networks since both the numbers of authors (nodes) and co-authorship links among them are growing over time. Additionally, the structure of the network (the way the authors are connected) and the positions of authors in the network may vary over time. Analysis of the attachment behaviour of authors (as nodes) in terms of the nodes' positional properties may help to explore the dynamics of structural change and evolutionary behaviour in scientific collaboration networks. Scopus[1] is one of the main widely used and authorized sources (in addition to ISI Web of Knowledge[2]) which store bibliometric data. It has an advanced search engine to look for publications based on a variety of combinations of fields. In the search results, it shows the number citations each publication has received. To construct our database, we extract publications which "information science" phrase has been used in their titles or keywords or abstracts by restricting to only the journal publications written in English. Indeed, the publications extracted cannot be considered as representing the world production in the field of "information science" but it illustrates a good portion of publications in this field that do not have limitation to a specific sub-field, conference, journal, institutions and country.

After extracting the publications meta-data, we used an application program, AcaSoNet, (Abbasi & Altmann, 2010) for extracting publication information (i.e. title, publication date, author names, affiliations, publisher, number of citations, etc.). It also extracts relationships (e.g., co-authorships) between authors and finally stores the data in the format of tables in its local database. Different types of information were extracted from each publication meta-data: publications information (i.e. title, publication date, journal name, etc.); and authors' names and affiliations (including country and institution names). As we are interested in different macro and micro level analysis (i.e. country and institution), affiliation data is so important for our research. Exploring our originally extracted data, we found affiliation information messy, as there were several fields missing for some of the publications and also differently written names for the country of origin and institutions. Thus, in our second step, we did some manual checking (using Google) to fill some missing fields (e.g., country) using other existing fields mainly the institution's name. We also tried to manually merge the institutions which had different written formats in our originally extracted dataset. After the cleansing of the publication data, the resulting database contained 4,579 publications reflecting the contributions of 10,130 authors and 22,962 co-authorships from 3,196 different institutions (i.e. universities and private companies) from 99 countries. All publications together received 49,154 citations.

---

[1] www.scopus.com.

[2] apps.isiknowledge.com

*Disaster Coordination Network Data*

During emergency and disaster situations individuals from different agencies cooperate to properly respond to the incident collectively. Inevitably, participants need to interact, communicate and cooperate with each other through the use of sharing information and experience, reporting and briefings, requesting resources and so on. Therefore, coordination network shapes, including agencies or participants from different organizations (agencies) as actors and their communication (interaction), represent the links or ties among actors in order to respond to the emergency. Coordinating activities or tasks in a complex system during a disaster for effective response is one of the most important issues to protect human, natural lives as well as from infrastructure damage. Therefore, a deep understanding of inter-organizational response network structure and the process of locating information flow and exchange is necessary in optimizing the response networks. This helps emergency managers and policy makers in making better informed decisions.

In 2009, the State of Victoria (Australia) experienced severe bush fires during February 2009. Although several agencies warned of the continuing fire threat and the forecast of extreme conditions for 6 and 7 February but "Conditions prevailing in the State on 7 February were unparalleled in Australia's history" (Clelland, Livermore, & Button, 2009). The 2009 Victorian

Bushfires Royal Commission was established on 16 February 2009 to investigate the causes and responses to the bushfires which swept through parts of Victoria in late January and February 2009. The data used in this research comes from content analysis by reviewing Victorian Bushfires Royal Commission situation reports after the February 7, 2009 (Black Saturday). Royal Commission situation reports were made available to the public and are retrievable from [http://www.royalcommission.vic.gov.au](http://www.royalcommission.vic.gov.au). The Kilmore East Fire (which was actually a group of separate fires) started on February 7. The Kilmore East Fire that ignited on 7 February 2009 was extraordinary and unprecedented. It was the fire with the most severe damages and fatalities. In all Kilmore East Fires, "119 people died and 1,242 homes were destroyed. The combined area burnt by the Murrindindi and Kilmore East fires, which later merged, was 168,542 hectares. The Kilmore East Fire alone burnt 125,383 hectares" (Government, 2009).

During the data collection process, we first reviewed the brief reports on the Kilmore East Fire. Then, identifying key participants mainly in the Incident Management Teams (IMTs) such as Incident Controller (IC) and its deputy (DIC), Planning Officers (PO), etc. involved in the emergency response network, we looked for these individuals' statements in the Royal Commission dataset. From the result of the content analyses of the 2009 Kilmore East Fire, we identified the main key actors involved in the response, and the support management mainly in the incident management team. Then, the statements of each key actor were extracted again searching the Royal Commission dataset. Applying content analyses on the key players' statements, their interactions were documented. Finally, 104 distinct actors and 286 interactions among them were extracted.

We have used UCINET (Borgatti, Everett, & Freeman, 2002), a social network analysis software program, to analyze the coordination network and its actors. UCINET is a comprehensive, well-accepted and long-standing program for the analysis of social networks. The program contains several network analysis routines (e.g., centrality measures, cohesion measures, positional analysis algorithms, and cliques), and general statistical and multivariate analysis tools (Borgatti et al., 2002).

**Testing and Evaluation**

Preliminary analysis of the data related to its distribution and potential relationships amongst variables needs to be considered. This can be done using descriptive statistics, histograms, normality tests and scatter plots. For normally distributed data, statistical tests examining relationships amongst variables such as Pearson's correlations can be used. However, if the data distribution is not normal, then non-parametric tests such as Spearman's rank correlation test should be applied. In order to answer the research hypotheses, the collaboration network data should be transferred to a matrix format such that we can apply mathematical analysis and metrics. Therefore, the social network analysis measures (e.g., individual centrality measures, network centralizations, network density, and network clustering coefficient) can be used in order to quantify individuals (actors) and networks properties and structural attributes.

*Characteristics of the Collaboration Networks Data*

Figure 2 below depicts the number of actors (authors) and their respective collaborations' frequency for academic collaboration networks (e.g., co-authorship networks of the Information Science research community) between 2001 and 2010. Therefore, the academic collaboration network can be regarded as a complex network in which the number of nodes and links in the networks is growing rapidly over time. This may lead to rapid change of the network structure. Figure 3 llustrates an increasing trend of collaborations count, which is much faster than the increment of the number of actors especially during the last periods. This highlights the existence of redundant links among already connected actors during the evolution of the collaboration networks.

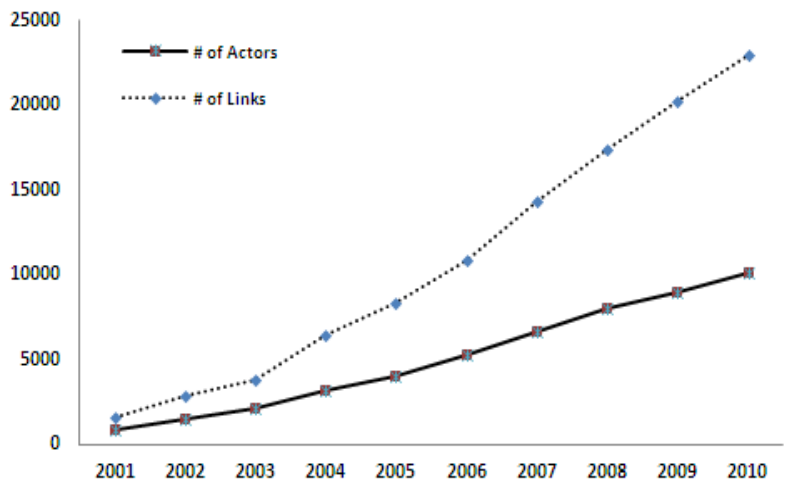

**Figure 2: Academic collaboration network's actors and collaboration cumulative frequency over time**

Figure 3 presents the number of actors and their collaborations' frequency for the Kilmore Fire disaster collaboration networks evolution over time. It shows that the disaster collaboration network is a complex network with the number of nodes and links in the networks growing over time, which leads to a change in the network structure. The disaster collaboration network growth over time shows an increasing trend of collaborations count, which is much faster than the increment of number of actors especially during the last periods. This highlights the existence of redundant links among already connected actors during the evolution of the collaboration networks.

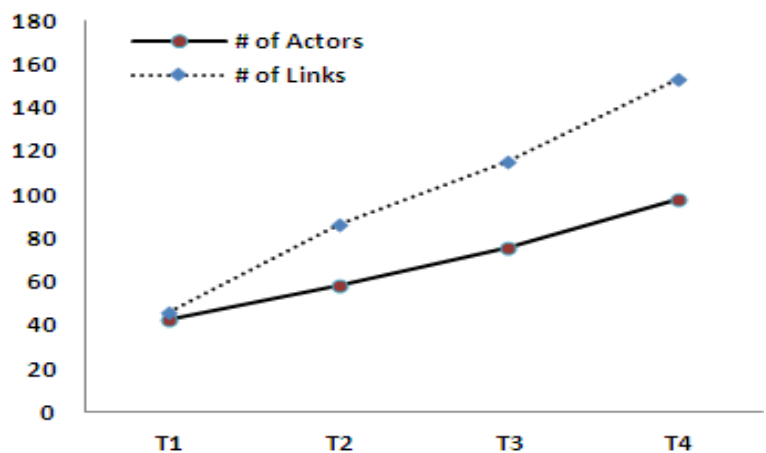

**Figure 3: Disaster collaboration networks actors and collaboration cumulative frequency over time**

*Small-World Properties of the Collaboration Networks*

It has been discussed earlier that most of the real-worlds networks have common properties (i.e. low density, high clustering, small diameter, and scale-free structure), which have also been called 'small-word properties' (Watts & Strogatz, 1998). In order to test the common properties of academic collaboration networks, these properties have been measured over ten years between 2001 and 2010. Cumulative numbers of authors (nodes) and links over time are used. For instance, the 2005 academic collaboration network statistics reported in Table 1 consist of all the authors (and their links) of the papers published between 2001 and 2005. Table 1 indicates the small-world property measures for each year. The results show that the academic collaboration network's *density* is very low for each period and is decreasing over time from 0.5% in 2001 to 0% in 2006, which remain unchanged over time. In addition, the results indicate that the academic collaboration network's *clustering coefficient* is high and remains almost constant with very small fluctuations between minimum of 0.72 at 2003 and maximum of 0.79 in the first year.

**Table 1: Academic collaboration network statistics and measures over time**

|  | 2001 | 2002 | 2003 | 2004 | 2005 | 2006 | 2007 | 2008 | 2009 | 20010 |
|---|---|---|---|---|---|---|---|---|---|---|
| **# of Actors** | 818 | 1466 | 2168 | 3220 | 4005 | 5320 | 6623 | 7992 | 9021 | 10130 |
| **# of Links** | 1571 | 2859 | 3792 | 6409 | 8307 | 10852 | 14334 | 17435 | 20259 | 22962 |
| **Sum of Links** | 1580 | 2903 | 3849 | 6513 | 8476 | 11040 | 14568 | 17735 | 20985 | 23730 |
| **Density (%)** | 0.5 | 0.3 | 0.1 | 0.1 | 0.1 | 0 | 0 | 0 | 0 | 0 |
| **Clustering Coefficient** | .79 | .75 | .72 | .74 | .74 | .74 | .74 | .75 | .76 | .76 |
| **Diameter** | 3 | 3 | 3 | 3 | 3 | 4 | 5 | 8 | 8 | 9 |
| **Average Distance** | 1.08 | 1.06 | 1.08 | 1.10 | 1.10 | 1.16 | 1.21 | 1.32 | 1.33 | 1.44 |
| **Degree Dist. Power-law $\lambda$** | 1.53 | 1.42 | 1.79 | 1.70 | 1.75 | 1.83 | 1.78 | 2.04 | 1.89 | 2.06 |

To examine the network diameter, the distance between each pair of nodes in the giant component (the largest connected component) has been explored. As shown in Table 1, the academic collaboration network's *diameter* is also very low but increases slowly over time (from 3 in 2001 to 9 in 2010) as the network grows. To consider a much-related concept, average distance between all pairs of nodes is measured. The result indicates a very low distance among each pair of nodes that is increasing slightly over time (from 1.08 in 2001 to 1.44 in 2010). Furthermore, the nodes degree distributions are also calculated over time. Figure 4 depicts the academic collaboration networks' degree distributions for four points in time (i.e. 2000, 2005, 2007, and 2010). The result of degree distributions implies that there are a few nodes with high degree but so many with low-degree nodes. Thus, the degree distributions follow a power-law distribution especially when the networks are larger (having more nodes and links).

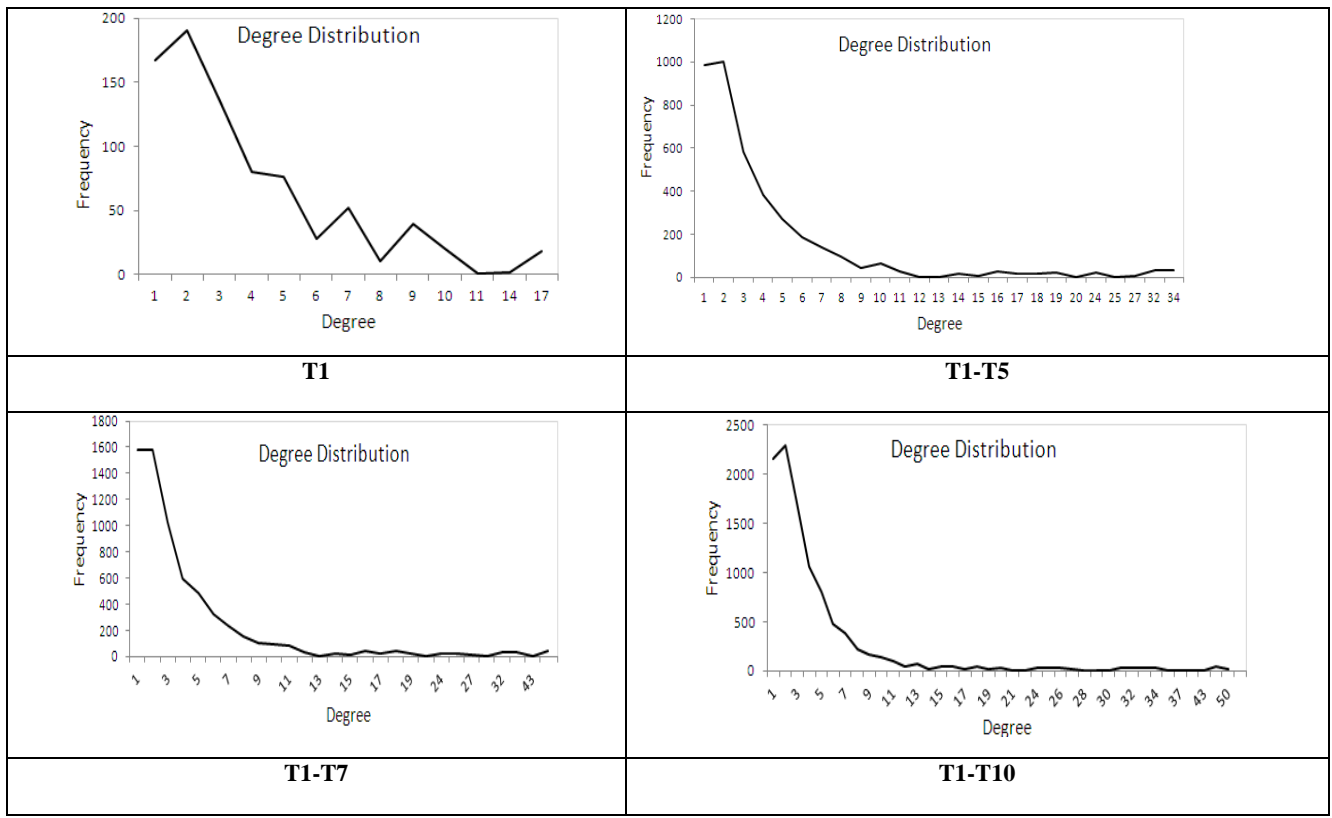

**Figure 4: Academic collaboration network's degree distribution**

In order to find the power-law coefficient (λ) of the network, the log-log distribution of degree and frequency for each year is drawn. Figure 5 depicts the academic collaboration networks' log-log degree distributions of the same years depicted in Figure 4, including the power-law coefficient and the explained variance $R^2$. The $R^2$ shows to what extent the linear function (the straight line) is fitted, the higher the $R^2$ values (i.e. those closer to one) the fitter the model. The power-law coefficient of the academic collaboration network is calculated for all years and shown in Table 1. The result implies that it is increasing over time (by growth of the network) from 1.43 in 2001 to 2.06 in 2010 (1.43 < λ < 2.06), with a few small fluctuations, respectively.

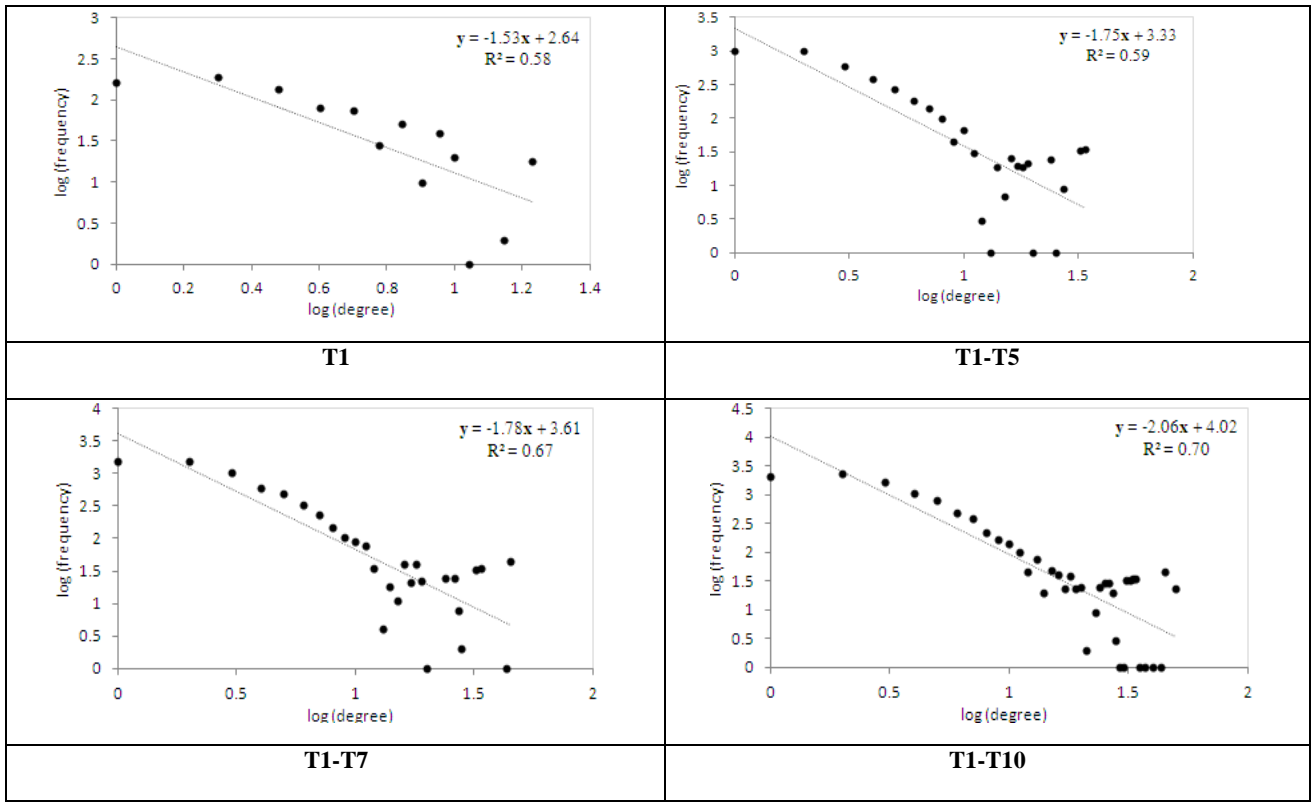

Figure 5: Academic collaboration network's log-log degree distribution over time

In conclusion, the academic collaboration networks of the Information Science research community has all the properties of small-world networks and can be considered a real small-world network with the result of its structural analysis maybe being extendable for other real-networks who share the common grounds of small-world properties.

**Disaster Collaboration Networks' Small-World Properties**

Similar to academic collaboration networks, the common properties of disaster collaboration networks (i.e. Kilmore Fire collaboration network) have been tested by measuring the properties of the disaster collaboration network. During the first day of the Kilmore Fire incident (February 7, 2009), 286 interactions among 104 distinct actors, individual people indicating the inter-personal cooperation (collaboration) network in response to the incident, is extracted.

The extracted interactions are mainly the ones for the day of the incident (7 February 2009) up to midnight in addition to some interaction prior to the incident day, during the preparedness phase, as the fire was forecasted due to hot weather and lack of humidity. To transfer the interaction data to longitudinal data for investigating the evolution of the network over time and to evaluate dynamic structural changes of the network, four important time points are used to make four consecutive periods (durations) expressing the evolution of the network during the disaster. The first point in time is at 11:50 (7 February) that the fire ignited. So, the first period (duration) contains any day before starting the fire (preparedness phase). The second point in time is 13:05 (7 February), when the Kilmore Fire station became the Incident Control Center (ICC). The third point in time is about 16:00 (7 February) that the Kilmore Incident Controller (IC) was replaced with a new IC. The last point in time is around mid-night (00:00, 8 February) as most of the logs and reports were available up to midnight). Table 2 shows the Kilmore Fire disaster interpersonal collaboration network statistics over time including a distinct number of actors (nodes), the number of interactions (number of links) and network properties (density, clustering coefficients, diameter, average path, degree distribution's power-law coefficient average degree and network centralization measures) for each period.

**Table 2: Disaster collaboration network statistics and measures**

|  | T1 | T1-T2 | T1-T3 | T1-T4 |
|---|---|---|---|---|
| # of Actors | 43 | 58 | 76 | 98 |
| # of Links | 46 | 86 | 115 | 153 |
| Sum of Links | 73 | 153 | 213 | 286 |
| Density (%) | 8.08 | 9.26 | 7.47 | 6.02 |
| Clustering Coefficient | .07 | .20 | .17 | .17 |
| Diameter | 5 | 5 | 5 | 5 |
| Average Distance | 2.60 | 2.77 | 2.94 | 2.93 |
| Degree Dist. Power-law $\lambda$ | 1.12 | 1.17 | 1.07 | 1.11 |

The results show that the disaster collaboration network's *density* is very low for each period. This means that the interpersonal response networks in all durations are rather sparse. The second duration network has the densest structure but the density of the network structure is decreasing over time.

In addition, these results indicate that the disaster collaboration network *clustering coefficient* is also low but contrary to network *density*, which shows increase over time from 0.07 to 0.17. This demonstrates the lack of many clusters (group of highly connected actors) in the disaster collaboration network and its growth during the response phase to the incident. As shown in Table 2, the disaster collaboration network's *diameter* is also very low and fixed at five for all periods. Moreover, average distance between all pairs of nodes is measured. The results show a very low distance among each pair of nodes (on average) which is increasing very slightly over time from 2.60 at T1 to 2.93 at T1-T4. Figure 6 depicts the disaster collaboration networks' degree distributions of four time periods. As shown, disaster collaboration networks follow a power-law distribution especially when the networks are larger. This means that there are few actors (nodes) having many connections (high degree) and there are many low-degree actors.

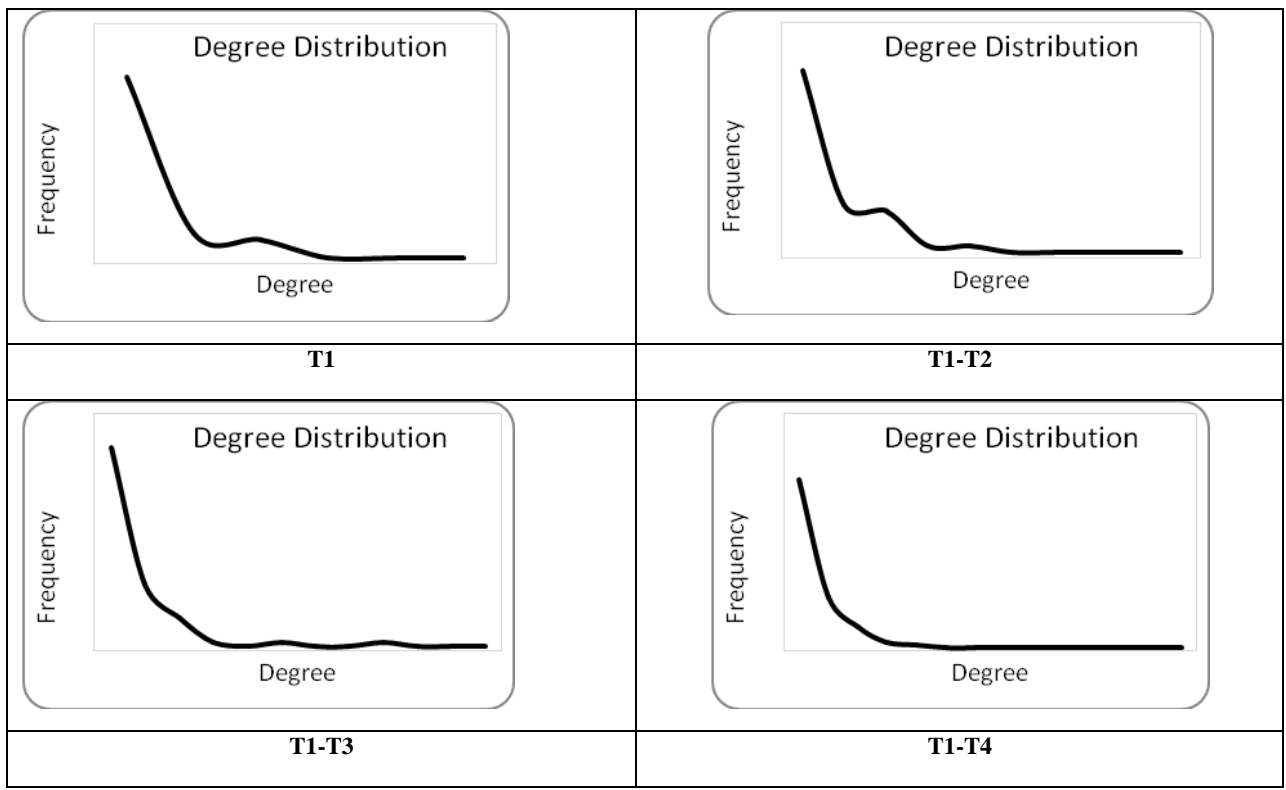

**Figure 6: Kilmore Fire disaster collaboration network degree distribution over time**

The log-log distribution of degree and frequency of each network is drawn for each period to calculate the power-law coefficient of the networks' degree distribution. Figure 7 depicts the disaster collaboration networks' log-log degree distributions of four time periods. The power-law coefficients of all periods are also reported in Table 2. The results indicate that the disaster collaboration degree distribution power-law coefficients show a moderate increase for the second period (T1-T2) and then decreasing over time during the third and fourth period. The disaster collaboration power-law coefficient is between 1.07 and 1.17 ($1.07 < \lambda < 1.17$).

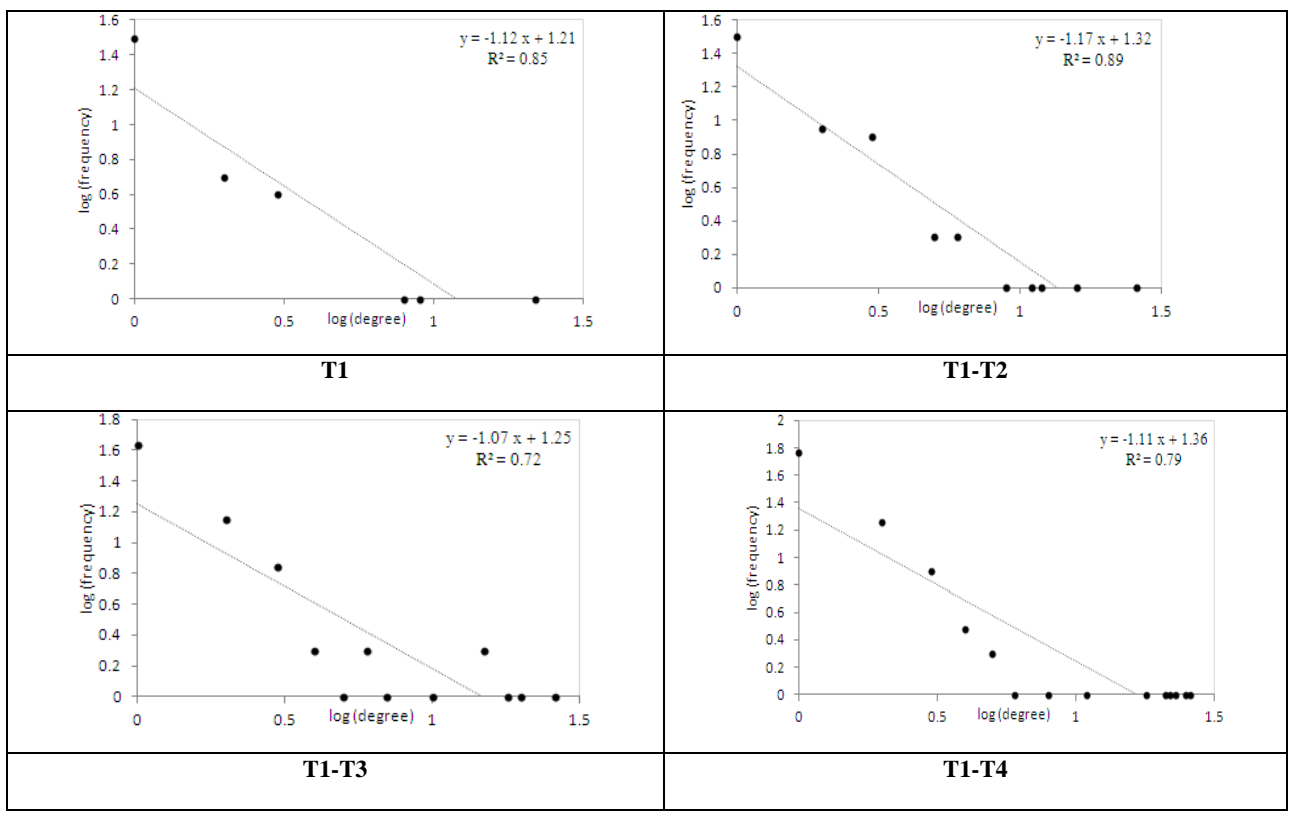

Figure 7: Kilmore Fire disaster collaboration network log-log degree distribution

In conclusion, the results demonstrate that the Kilmore Fire disaster collaboration networks represent a very sparse scale-free network structure with low diameter, although their clustering coefficient is very low. Therefore, based on the literature, the Kilmore Fire disaster collaboration network cannot be considered as a small-world network and the results for analyzing its structure and evolution may not be applicable in other cases.

**Conclusions**

This study synthesized different theories and approaches on the evolution and adaptation of complex networks from various fields for exploring the attachment behavior of actors in the networks. The adaptation behavior of complex networks over time in spite of dynamic changes of network structure, due to entrance of new actors and links to the network over time, are also examined. First, the common properties of small-world networks are examined on the two sample collaboration networks used in this study. The results show that the academic collaboration network (i.e. co-authorship networks of scholars in the field of Information Science) is very sparse and highly clustered with low diameter and actors' degree distribution has a scale-free structure. Therefore, the results of its networks analysis may be applicable to other complex networks which have similar common properties (Watts & Strogatz 1998). But the Kilmore Fire disaster collaboration network, as a representation of a network in a dynamic and complex environment, cannot be considered as small-world networks as its clustering is low.

Then, in light of the analyzing mechanism of (network) evolution (i.e. selection, retention and variation), the attachment behavior of actors are investigated to find whether attributes of actors attract more actors to attach to them. In other words, the attributes of actors which impact on the fitness function that other actors use to select them for attachment (in the selection process) is evaluated. To do this, actors' different logics of attachment are examined on two sample collaboration networks. The results of the longitudinal analysis supported the existence of at least three of them for both collaboration networks. Multi-connectivity (having divers partners), homophily (having partners with similar connectivity) and preferential attachment (having tendency to authors who have high betweenness centrality (Abbasi, Hossain, & Leydesdorff, 2012)) have been found having impact on the link formation (attachment) of authors in the academic collaboration network over time. On the other hand, multi-connectivity, preferential attachment and embedding (weakly) found to have effects on actors' attachment in the disaster collaboration network. Then to investigate the retention mechanism of the networks, the dynamic of networks is tested to see how the topology of the network changes in time according to actors' attachment logics. Actors' log-log degree distribution's exponent (λ), Pearson degree correlation, average strength, and actors' average neighbors' degree are used as proxies for preferential attachment, homophily, embedding and multi-connectivity, respectively.

Correlation tests between the networks topology (measured by network centralization) and the actors' attachment logics are applied to identify which attachment logic leads to higher network topology change. The result of the correlation test for the academic collaboration network indicates actors' preferential attachment as the most influential driver of network topology changes. But multi-connectivity is detected as the most effective driver of network topology changes in the disaster collaboration network although other logics of attachment were found effective as well for both networks. Finally, to evaluate the variation mechanism of networks, the dynamic on networks is examined to see what attributes of the network topology remains static while the states of the actors change over time. The results illustrate that the networks' clustering coefficients remain almost static over the evolution of both networks while the network is expanding over time through adding new actors and links to the networks.